\documentclass[conference,a4paper]{IEEEtran}
\IEEEoverridecommandlockouts
\usepackage{cite}
\usepackage{amsmath,amssymb,amsfonts}
\usepackage{algorithmic}
\usepackage{graphicx}
\usepackage{textcomp}
\usepackage{xcolor}
\usepackage{comment}
\def\BibTeX{{\rm B\kern-.05em{\sc i\kern-.025em b}\kern-.08em
    T\kern-.1667em\lower.7ex\hbox{E}\kern-.125emX}}

\begin{document}
\bstctlcite{IEEEexample:BSTcontrol}	

\title{A Multidimensional Elasticity Framework for Adaptive Data Analytics Management in the Computing Continuum\\

}

 \author{Sergio Laso\IEEEauthorrefmark{1}, Ilir Murturi\IEEEauthorrefmark{2}, Pantelis Frangoudis\IEEEauthorrefmark{2}, Juan Luis Herrera\IEEEauthorrefmark{1}, Juan M. Murillo\IEEEauthorrefmark{1}, and Schahram Dustdar\IEEEauthorrefmark{2}  
 \IEEEauthorblockA{\\\IEEEauthorrefmark{1}{Universidad de Extremadura}, Cáceres 10003, Spain. e-mail: slasom@unex.es}
  \IEEEauthorrefmark{2}{Distributed Systems Group}, {TU Wien}, Vienna, 1040, Austria.
 }

\maketitle

\begin{abstract} 
The increasing complexity of IoT applications and the continuous growth in data generated by connected devices have led to significant challenges in managing resources and meeting performance requirements in computing continuum architectures. Traditional cloud solutions struggle to handle the dynamic nature of these environments, where both infrastructure demands and data analytics requirements can fluctuate rapidly. As a result, there is a need for more adaptable and intelligent resource management solutions that can respond to these changes in real-time. This paper introduces a framework based on multi-dimensional elasticity, which enables the adaptive management of both infrastructure resources and data analytics requirements. The framework leverages an orchestrator capable of dynamically adjusting architecture resources such as CPU, memory, or bandwidth and modulating data analytics requirements, including coverage, sample, and freshness. The framework has been evaluated, demonstrating the impact of varying data analytics requirements on system performance and the orchestrator's effectiveness in maintaining a balanced and optimized system, ensuring efficient operation across edge and head nodes.
\end{abstract}


\section{Introduction} \label{Sec:introduction}

The exponential growth of data, driven by the widespread adoption of IoT devices, has transformed sectors such as healthcare, industry, and smart cities \cite{gkonis2023survey, herrera2022joint}. This data explosion requires advanced data analytics for these sectors to generate actionable insight from vast, complex datasets. However, cloud infrastructures are increasingly struggling to meet the stringent quality, responsiveness, and real-time processing requirements of these applications. The complexity of interconnecting large numbers of devices, combined with the need to meet data analytics requirements and decision-making, has highlighted the limitations of traditional cloud solutions \cite{herrera2021optimizing}. In response to these challenges, the Computing Continuum (CC) has emerged as a solution integrating cloud, edge, and IoT environments, enabling more efficient resource distribution and real-time processing closer to the data source~\cite{bendechache2020simulating, casamayor2022distributed }.

Nevertheless, the management of data analytics requirements within an application can evolve over time, which can lead to sub-optimal performance, and it is crucial to dynamically adapt resources in response to changing demand. Incorporating \emph{elasticity} thus becomes essential \cite{pujol2023edge}. Elasticity~\cite{dustdar2011principles} can be defined along three fundamental dimensions: quality, resources, and costs. In previous work~\cite{laso2022elastic}, we explored quality-related elasticity for data analytics, but this approach is limited. In complex applications, elasticity must go beyond quality to encompass all dimensions. This can provide a more complete view of the system and enable more effective management, ensuring optimal performance in a CC context.

This work presents a multi-dimensional elasticity management approach designed to adaptively handle both infrastructure and data analytics requirements. Unlike traditional resource-level elasticity models, multi-dimensional elasticity further incorporates data analytics-related requirements; coverage, sample, and freshness. By accounting for both resource management and analytics needs, this approach provides a richer adaptation space that can effectively respond to varying data demands. When infrastructure resources can no longer scale, the system can adapt analytics requirements to maintain acceptable performance. This flexibility ensures that, even with limited resources, analytical tasks can be executed without compromising the operability of the infrastructure. In this direction, we make the following contributions:

\begin{itemize}
    \item A framework (\S\,\ref{Sec:architecture}) for multi-dimensional elasticity management in the CC, tailored to data analytics applications. It encompasses the three key data analytic requirements (\S\,\ref{sec:analytics-requirements}): coverage, sample, and freshness, alongside traditional elasticity dimensions \cite{murturi2022decent}, ensuring a balanced and adaptive approach to data analytics management.
    
    \item A Reinforcement Learning (RL)  component for managing resource elasticity in CC environments (\S\,\ref{sec:rl-predictor}). This RL-based predictor is integrated into an orchestrator (\S\,\ref{sec:orchestrator}) which enables dynamic and autonomous resource adaptation to meet analytics requirements while balancing cost and other resource constraints. 
    

    \item  An experimental evaluation (\S\,\ref{Sec:validation}) of the impact of data analytics requirements and their adaptations on compute resource consumption, as well as of the decision-making performance of our RL-based orchestrator in terms of meeting analytics quality, resource consumption, and infrastructure cost objectives. 


\end{itemize}

\section{Related Work} 
\label{Sec:background}

Management and orchestration in the CC has received significant attention.
In~\cite{ullah2023orchestration}, a comprehensive survey of orchestration strategies is presented, highlighting the various challenges and solutions for managing distributed computing resources. This study delves into the state-of-the-art orchestration frameworks and their use in hybrid cloud environments.

The integration of AI with orchestration strategies is discussed in \cite{wu2020cloud}. It emphasizes how AI techniques, especially machine learning, can optimize resource allocation, workload balancing, and dynamic adaptation in cloud continuum environments. Technology-oriented orchestration solutions are examined in \cite{vcilic2021towards}, where the authors discuss how modern technologies, such as those related with containerization, are employed to manage resources in edge-cloud infrastructures. This study highlights the practical deployment issues and potential technological advances required to support scalable orchestration. Although these surveys provide a rich understanding of orchestration frameworks, they often lack practical proposals to deal with the particularities of data analytics. 
On the simulation side, \cite{luckow2021pilot} focuses on simulation-based approaches for managing scenarios in the cloud continuum but lacks practical results for real-time resource management. Lastly, in \cite{masip2021managing}, the authors present insights into managing cloud-to-edge infrastructures but emphasize the need for a more holistic approach to address the complexity of modern IoT and cloud environments.

In conclusion, while many studies offer valuable insights into orchestration and resource management in cloud continuum environments, there is a clear gap in the practical proposal or implementation of multi-dimensional elasticity for real-time analytics, which this work aims to address.

\section{Adaptive Elasticity for Data Analytics in the CC} \label{Sec:architecture}



\begin{figure*}[ht]
\centering
\includegraphics[width=\textwidth]{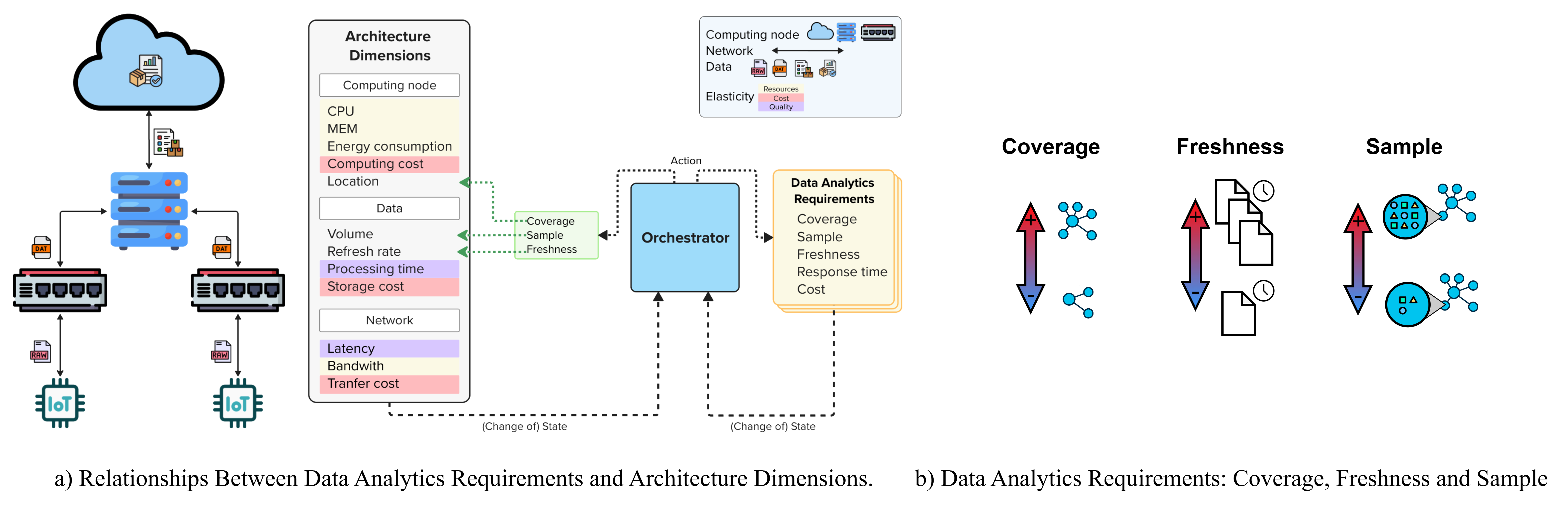}
\caption{Multidimensional Elasticity Management in CC}\label{fig:elasticityDimensions}
\vspace{-5mm}
\end{figure*}

Elasticity within the CC architecture is fundamental to enabling dynamic adaptation of both infrastructure resources and data analytics requirements. This elasticity spans multiple dimensions that impact both the operational efficiency and the performance of analytics applications. In this context, we propose a comprehensive framework for managing multi-dimensional elasticity in data analytics for the CC. These dimensions, illustrated in Fig.\ref{fig:elasticityDimensions}a, reflect the different factors that influence the adaptability of both computing resources and the data analytics themselves.

At the operational level, elasticity focuses on the efficient use of resources like CPU, memory, and energy, ensuring that these computing resources can be dynamically adjusted to changing workloads. This level also involves managing the cost associated with using infrastructure resources, particularly in environments where resource consumption is billed based on usage. At the application level, elasticity manages the specific requirements of data analytics, such as the quality of results, the frequency at which data is processed, and the timeliness of the analytics outcomes. These dimensions require careful balancing by the orchestrator to ensure that the infrastructure can meet performance expectations without exceeding resource or cost limitations. Furthermore, these dimensions are interrelated, since changes in one can affect others. E.g., increasing the sample size to improve the quality of analytics, leads to higher resource consumption, impacting the CPU, and bandwidth, and potentially increasing costs. 

To achieve this, we propose an RL-based predictor component integrated within the orchestrator. This RL-based predictor enables dynamic and autonomous resource allocation, ensuring that the system can meet the specific data analytics requirements (e.g., freshness, coverage, sample size, response time) while considering cost and resource constraints. However, if the architecture cannot provide the necessary resources, the orchestrator can also modify the initial data analytics requirements within predefined boundaries.


\subsection{Data Analytic Requirements}
\label{sec:analytics-requirements}
Data analytics requirements define the key parameters for executing analytics tasks and are dynamically managed by the orchestrator as part of the elasticity framework. These requirements are defined within certain limits and can be encoded as Service Level Objectives (SLOs), allowing the orchestrator to modify them as necessary to optimize the balance between system performance and resource availability. The core data analytics requirements managed by the orchestrator include:

\begin{itemize}
    \item Freshness: Indicates the frequency at which the analytics are to be executed. Depending on the purpose of the analytics, it may be necessary to execute them more or less frequently within a given period. 
    
    \item Coverage: Indicates the available locations from which analytical data is desired. It can be specified using identifiers provided by the orchestrator, which are associated with different locations in the infrastructure.
    
    \item Sample: This requirement indicates the percentage of representative samples desired from the specified locations. Within each location, various devices are generating the required information. Depending on the analytics' objective, a complete representative sample or only a portion can be chosen using this parameter.
    
    \item Response Time: Sets the response time desired for delivering the requested analytics. 
   
    \item Cost: Represents the desired budget available to pay for the execution of the analytics. 

\end{itemize}

Finally, Figure \ref{fig:elasticityDimensions}b shows the graphical impact of the coverage, freshness, and sample requirements. Expanding the coverage area increases the number of nodes involved in the analysis while increasing freshness requires more frequent updates and data processing. Similarly, a larger sample size means more data needs to be processed and transferred.


\subsection{Architecture Dimensions}
\label{sec:architecture-dimensions}
Dimensions are grouped into three categories based on the different entities in the architecture: computing nodes, networks, and data. Each category encapsulates critical dimensions that the orchestrator considers when adapting the system to meet analytics requirements and maintain performance.

\begin{itemize}
    \item Computing node: Represents the hardware responsible for executing data analytics within the infrastructure. Key metrics monitored include \textit{CPU} usage, which tracks processing capacity, available RAM (\textit{MEM}), geographical \textit{location}, which identifies where the node is situated, \textit{energy consumption}, referring to the power usage, and \textit{computing cost}, indicating the expense associated with utilizing the node's computational resources.


    \item Network: The network entity encompasses various aspects related to data transfer and communication between computing nodes. Key metrics include \textit{bandwidth}, which refers to the available capacity for data transfer, \textit{transfer cost}, which is the expense associated with moving data across the network, and \textit{latency}, which measures the time taken to transfer data between nodes.
    \item Data: The data entity focuses on the properties of the data being processed within the infrastructure. Key metrics include \textit{volume}, which denotes the total amount of data, providing insight into the scale of data processed for resource optimization. \textit{Refresh rate} refers to the frequency at which data is updated from its source, indicating how often new data is available. \textit{Processing time} measures the duration between the start of data analytics and the retrieval of results. Lastly, storage cost reflects the expense associated with storing data in the infrastructure.
            
            
            

\end{itemize}

\subsection{Relationships Between Data Analytics Requirements and Architecture Dimensions}
\label{sec:relationships}

Understanding the relationships and impacts that data analytics requirements have on various dimensions of the infrastructure is critical to managing the performance, resource allocation, and overall efficiency of the CC architecture.

\noindent\textbf{Direct relationships} refer to explicit connections between specific data analytics requirements and architecture dimensions. These relationships show how a change in one element directly influences the behaviour or performance of another.

\begin{itemize}
    \item \textit{Coverage and Location:} Coverage determines the geographical locations from which data analytics are conducted, creating a direct link with the Location of computing nodes. A broader coverage area requires more geographically dispersed computing nodes to collect and process data from a larger set of locations.
    
    \item \textit{Sample and Volume:} The Sample size refers to the percentage of data or devices used for data analytics. A larger sample size results in more end devices or data points contributing to the analysis, which in turn increases the total volume of data to be processed and stored.
    
    \item \textit{Freshness and Refresh rate:} Freshness measures how up-to-date the data is, and it is directly tied to the refresh rate of data updates. A higher refresh rate ensures more frequent updates, thereby maintaining the real-time relevance and freshness of the data being analyzed.


\end{itemize}

\noindent\textbf{Indirect relationships} are more complex and less apparent connections between data analytics requirements and infrastructure dimensions, but they can significantly impact system performance. This work focuses specifically on the impact of coverage, sample, and freshness for CPU. Larger coverage and sample size entail processing data from more computing nodes and data sources, thus increasing CPU usage. Lastly, higher freshness involves more frequent data updates to process, which places a greater load on CPU resources. These indirect effects on CPU usage are critical in shaping overall system performance, as highlighted in Section~\ref{Sec:validation}.

\subsection{Orchestrator}
\label{sec:orchestrator}

\begin{figure}[ht]
    \centering
    \includegraphics[width=0.5\textwidth]{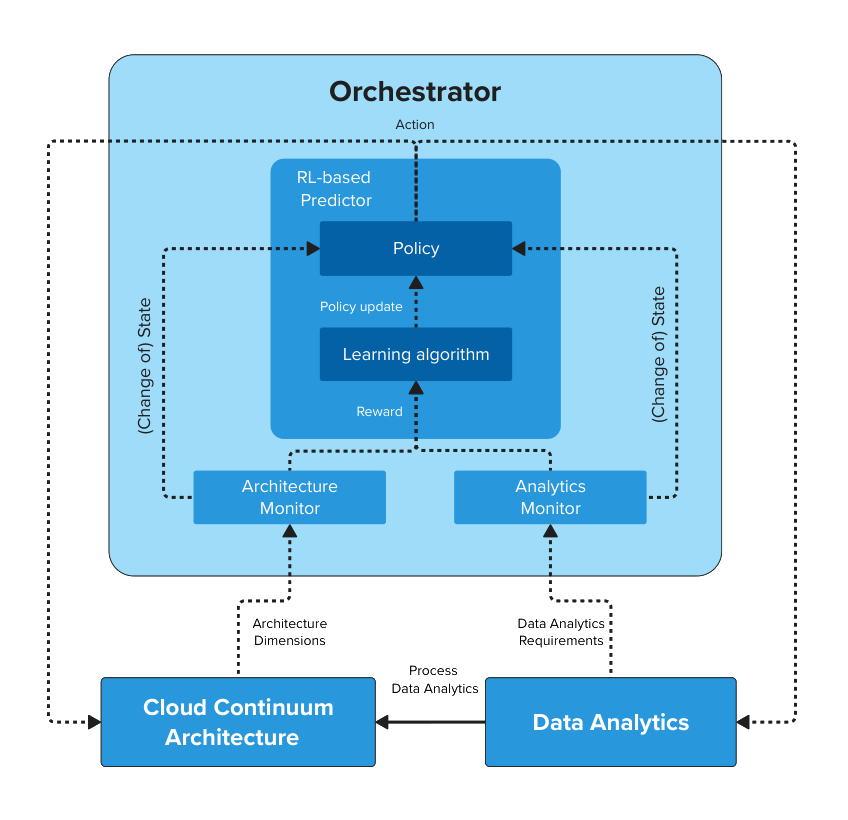}
    \caption{Orchestrator with RL-Based Elasticity Management in the CC.}\label{fig:orchestrator}
\end{figure}

The orchestrator is a pivotal component in the CC architecture, responsible for jointly managing the elastic requirements of data analytics and adapting the infrastructure to the evolving needs of both the infrastructure and the analytics. 
%
In our framework, the core decision-making logic of the orchestrator is implemented by an RL-based agent that carries out run-time adaptations at the resource and analytics levels. This two-level dynamic adaptation  ensures a balance between performance, cost, and resource availability.
The orchestrator consists of three main modules that work in tandem, as illustrated in Fig.~\ref{fig:orchestrator}, which are detailed below:

    \subsubsection{Architecture Monitor} This module is responsible for continuously tracking the resource dimensions of the architecture. It gathers metrics related to available resources such as CPU usage, memory, network bandwidth, or energy consumption. Additionally, it monitors costs and other critical data points, such as response time and service-level performance, ensuring that the orchestrator has a real-time snapshot of the system's operational status. 

    \subsubsection {Analytics Monitor} The Analytics Monitor oversees the status of the data analytics running on the system. This module collects and stores consumers' requirements, such as data freshness, coverage, and sample size. These requirements are then compared with the real-time status of the system, as reported by the Architecture Monitor, to identify potential mismatches or inefficiencies. 
    %

    \subsubsection{RL-based Predictor} 
    \label{sec:rl-predictor} 
    By interacting with the environment (CC and the deployed analytics application), this module learns how to take appropriate adaptation actions at runtime. The space of available actions is defined along the three elasticity dimensions, with the goal of meeting quality, cost, and resource-related requirements expressed as SLOs. In the absence of a precise model of the environment and of the way adaptation decisions affect SLO fulfillment, we apply a Q-Learning (QL) framework on which we elaborate below, to learn the optimal adaptation policy.
    
    \paragraph{System model} We design a QL agent that obtains system-wide state information from the Architecture Monitor (CPU utilization, capacity) and the Analytics monitor (analytics application configuration) and applies specific actions at discrete time steps. Each action brings the system to a new state and comes with a reward that the agent receives. The agent aims to derive a value function $Q: \mathcal{S} \times \mathcal{A} \rightarrow \mathbb{R}$, which represents the expected cumulative discounted future reward if the agent selects action $a \in \mathcal{A}$ at state $s \in \mathcal{S}$ and continues by following the optimal policy. 
    
    \paragraph{State ($\mathcal{S}$) and action ($\mathcal{A}$) spaces} A state encodes the following information: (i) node CPU utilization, (ii) the amount of resources allocated at compute nodes, and (iii) the analytics configuration in terms of freshness (data collection frequency) and sample size (percentage of data sources activated). At each state, the agent picks an action that encodes (i) the analytics application configuration and (ii) the resource configuration, i.e. the amount of compute capacity allocated to nodes. Note that these actions capture the quality (data freshness, sample size), resource (CPU utilization, allocated capacity), and cost (allocated capacity) elasticity dimensions.
    
    We make a number of simplifications. First, compute nodes and the respective analytics application instances are considered identical, i.e., the same amount of resources is allocated, the same processing capabilities are assumed, and the analytics configuration is the same. In consequence, node CPU utilization state is considered binary: if there is at least one compute node that is overloaded, the system as a whole is considered over-utilized. 

    Second, we discretize the state and action spaces.
    There is a fixed number of possible resource allocations, and scaling in/out is implemented by selecting the number of application instances to launch on a node out of a fixed (small) set of values; this would translate to spinning up a new container to add data processing capacity. The same applies to analytics application configurations. frequency and sample size are restricted to a small number of levels. For example, in our evaluation, each of the available resource and analytics configuration actions can take one of four values. Combined with the two potential CPU utilization states, this leads to a space of $2 \times 4 \times 4 \times 4$ states. 
    
    Finally, in this instantiation of the QL agent, we do not consider analytics coverage. Assuming identical nodes and application instances, the state and action spaces can be extended to account for this case, e.g. by adding a state component encoding the number of locations/nodes active and an action to drop (add) a random compute node to reduce overall CPU utilization (increase coverage).
    
	\paragraph{Reward structure} We assume SLOs are in place to encode infrastructure and application requirements. In particular, let $SLO_{CPU}$, $SLO_{c}$, $SLO_{f}$, and $SLO_{s}$ denote the required maximum CPU utilization (percentage), maximum cost in terms of the number of instances deployed, minimum data freshness (frequency) and minimum sample size (percentage). If any such SLO is met, this contributes a fixed amount (respectively, $r_{CPU}$, $r_{c}$, $r_{f}$, and $r_{s}$) to the collected reward. Otherwise, this fixed amount is subtracted. This design provides flexibility to prioritize different operator preferences by tuning these amounts. For example, in our evaluation, we set $r_f = 10$ and $r_{CPU} = r_{c} = r_{s} = 1$ to drive our QL agent towards actions that favor $SLO_f$ fulfillment, potentially at the expense of other SLO violations.

\section{Evaluation} \label{Sec:validation}
\subsection{Methodology}
In this section, we evaluate our elasticity framework following a twofold strategy. First, via testbed experiments, we demonstrate the impact of data analytics requirements and their adaptations on resource (CPU) usage. To do so, we apply our system to a realistic use case from the Smart Cities domain. Second, via simulation, we showcase the effectiveness of our QL agent to learn the appropriate adaptation policies and flexibly support different operator and customer priorities in terms of SLO fulfillment.

\subsection{Case study and experimental setup}
We present a case study involving a smart city deployment, where the city aims to improve urban planning and public safety through advanced data analytics. The system monitors human movement in different areas of the city to generate heatmaps that help identify patterns, risks, and opportunities for optimizing city services. The city has implemented a distributed infrastructure with nodes equipped to process and store data analytics. The orchestrator coordinates the execution of data analytics, ensuring that the specified data analytics requirements are met. The infrastructure is hierarchical, with a central head node aggregating data from edge nodes located across different parts of the city. These edge nodes collect information from local sensors monitoring a person's movement. 

We evaluate the impact of data analytics requirements (i.e., coverage, sample, and freshness) on CPU usage across the infrastructure. The infrastructure consists of one \textit{headnode} and three edge nodes (\textit{edgenode1}, \textit{edgenode2}, \textit{edgenode3}), simulating a system that generates heatmaps of the city's different areas using the One simulator to generate the person movement dataset~\cite{jesus2021oppnets}. Edge nodes generate heatmaps of their respective areas, which are then aggregated by the head node to produce a unified heatmap. 

\subsection{Impact of data analytics requirements}
\begin{figure*}[htpb]
\centering
\includegraphics[width=\textwidth]{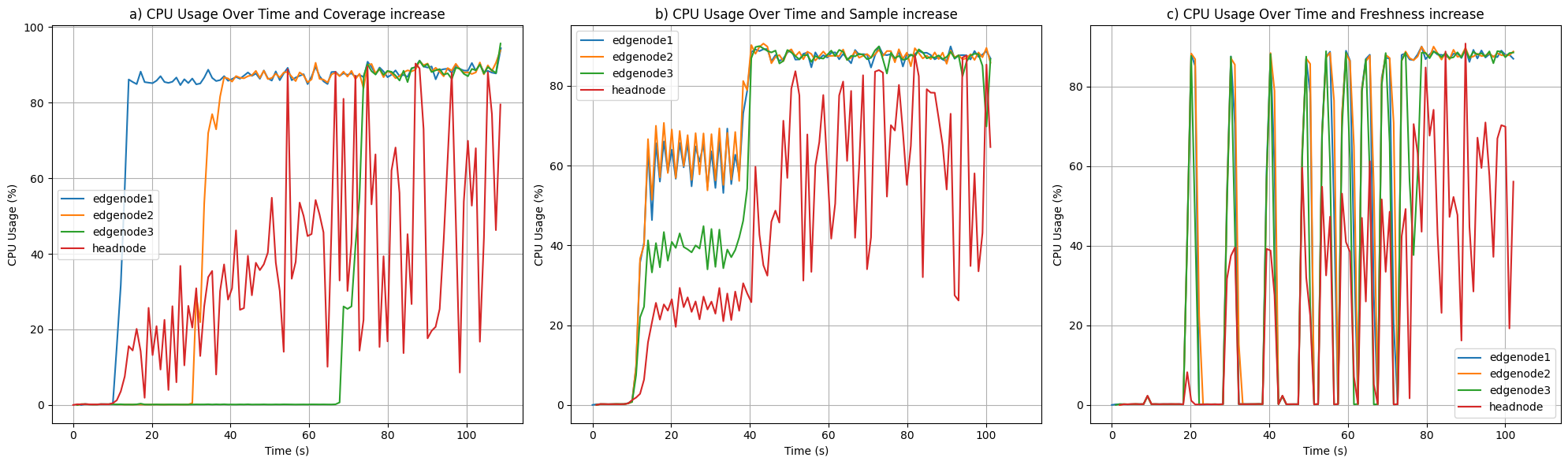}
\caption{Impact of data analytics requirements on CPU usage.}\label{fig:dataAnalyticsImpact}
\end{figure*}

In Fig.~\ref{fig:dataAnalyticsImpact}a, we observe that as coverage increases, more edge nodes are progressively engaged, leading to a staggered rise in CPU usage. Initially, only one \textit{edgenode} is active, but as more geographic areas are covered, additional nodes become involved, creating a stepwise pattern in CPU utilization. Simultaneously, the \textit{headnode's} CPU usage also rises as it processes the increasing amount of data from multiple edge nodes. In Fig.~\ref{fig:dataAnalyticsImpact}b, as the sample increases, CPU usage gradually rises across all edge nodes. Similarly, the \textit{headnode} shows a CPU increase as it processes the increasing volume of data. This behavior reflects how larger samples translate to the collection of more data for analytics, resulting in higher processing demands for both the edge nodes and the \textit{headnode}, which aggregates all the incoming data. In Fig.~\ref{fig:dataAnalyticsImpact}c, as freshness increases, there is a higher demand for frequent data updates. The edge nodes show a relatively constant CPU usage but with an increased periodicity of spikes as freshness increases. This indicates that the system is processing updates more frequently. However, the \textit{headnode} exhibits a more gradual increase in CPU usage, reflecting the higher frequency of data processing as it aggregates information from the three edge nodes at an increased rate. In summary, the assessment highlights the significant impact of these data analytics requirements on the entire CC infrastructure. These results underscore the importance of multi-dimensional elasticity and the ability of the orchestrator to dynamically adjust these analytics requirements in the face of resource scaling limitations, ensuring a flexible and balanced response to fluctuations in workloads and resource availability.

\subsection{Impact of adaptation actions}
\begin{figure}[ht]
    \centering
    \includegraphics[width=0.4\textwidth]{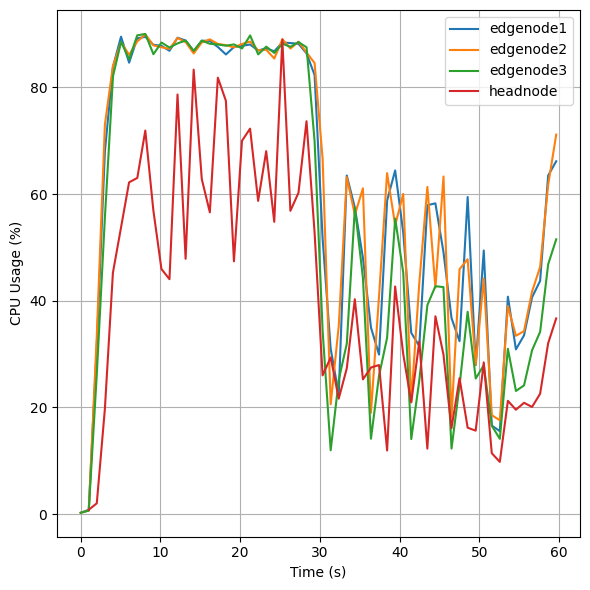}
    \caption{Orchestrator adapting data analytics requirements to manage CPU usage.}\label{fig:orchestration-analytics}
    \vspace{-4mm}
\end{figure}

We then present an experiment that demonstrates the system's behavior at the resource level when adaptation decisions are made. The orchestrator monitors CPU utilization across all infrastructure and adjusts the data analytics requirements dynamically to prevent overloading any of the nodes. 
In the example of Fig.~\ref{fig:orchestration-analytics}, it can be observed how the CPU shows an increase, prompting the orchestrator to pick an action that adjusts the sample size to reduce the load. The system then moves to a state where high usage persists. The next orchestrator action is to modify freshness, decreasing the update frequency, thus transitioning to a state with low CPU usage and maintaining system balance.


\subsection{RL-based predictor performance}
We finally demonstrate the decision-making performance of our QL-based predictor. We simulate an environment with four nodes and configure them so that they have the same average data processing capacity and the same average number of data sources (e.g., IoT devices) attached to them. At each time step, the QL agent picks one of the admissible system configurations, each prescribing a combination of freshness (update frequency), sample size, and node capacity (and thus deployment cost). There are four distinct possible values for each of these parameters, leading to a space of 64 actions, while there are 128 states. The CPU utilization SLO is set to $SLO_{CPU} \leq 80\%$, while analytics quality SLOs are set such that $SLO_f$ is at least half of the maximum possible data collection frequency, and $SLO_{sample} \geq 50\%$ of the data sources. The cost-related $SLO_c$ is defined relative to the maximum allowed per-node capacity; the more the capacity, the more the cost. To elaborate, there are four scaling levels with increasing infrastructure cost, and node resources should not be scaled to more than $SLO_c$, otherwise this SLO is violated. If under a given configuration the data intensity (volume of data to process per second) is higher than $SLO_{CPU}$ of the node's nominal data processing capacity, then with high probability the node transitions to an overloaded state. We run five sets of experiments, each with a different reward configuration and thus performance priority, and measure the SLO fulfillment ratio after the QL algorithm has converged. As Table~\ref{tbl:slos} shows, when the relative importance of a specific SLO is high (e.g., when freshness matters more, the $r_f$ reward component is 10$\times$ higher than the other components), the QL-based predictor fulfills this SLO close to 100\% of the time, while other SLOs may suffer. Putting equal importance to all SLOs, on the other hand, achieves a good trade-off (last row of Table~\ref{tbl:slos}).

\begin{table}
	\caption{SLO fulfillment ratios for different reward configurations.}
	\label{tbl:slos}
		\centering
		\setlength\tabcolsep{4pt}
		\begin{tabular}{|l|c||c|c|c|c|}
			\hline
			Priority & Reward structure & \multicolumn{4}{|c|}{SLO fulfillment ratio}\\
            \cline{3-6}
            &($r_{CPU}:r_f:r_s:r_c$) & CPU load &  Freshness & Sample  & Cost\\
			\hline
			CPU load & \textbf{10}:1:1:1 & \textbf{0.996} & 0.920 & 0.795 & 0.797\\
			Freshness & 1:\textbf{10}:1:1 & 0.894 & \textbf{0.992} & 0.804 & 0.809\\
			Sample & 1:1:\textbf{10}:1 & 0.873 & 0.913 & \textbf{0.984} & 0.781\\
			Cost & 1:1:1:\textbf{10} & 0.875 & 0.908 & 0.783 & \textbf{0.984}\\
			Balanced & 1:1:1:1 & 0.974 & 0.979 & 0.953 & 0.952\\
			\hline	
		\end{tabular}
\end{table}

\section{Conclusion} \label{Sec:conclusion}

 Elasticity within the CC is critical in enabling infrastructure and application requirements to adapt seamlessly and responsively to changing conditions. Therefore, this paper presented a novel framework for managing elasticity in the CC, focusing on the dynamic needs of data analytics. 
 Integrating RL concepts showed promising results in terms of jointly and flexibly adapting infrastructure resources and analytics configurations. This opens up interesting directions for future work, particularly towards devising more sophisticated RL strategies and further operationalizing our framework.


%

\section*{Acknowledgment}

This work was partially funded by grant DIN2020-011586, projects TED2021-130913B-I00, PDC2022-133465-I00 (MCIN/AEI/10.13039/501100011033 and EU NextGenerationEU/PRTR), the Junta de Extremadura (GR21133), the European Regional Development Fund, and the EU Horizon Europe program under grants 101135576  and 101079214.


\bibliographystyle{IEEEtran}
\bibliography{IEEEabrv,mybib}

\begin{thebibliography}{10}
\providecommand{\url}[1]{#1}
\csname url@samestyle\endcsname
\providecommand{\newblock}{\relax}
\providecommand{\bibinfo}[2]{#2}
\providecommand{\BIBentrySTDinterwordspacing}{\spaceskip=0pt\relax}
\providecommand{\BIBentryALTinterwordstretchfactor}{4}
\providecommand{\BIBentryALTinterwordspacing}{\spaceskip=\fontdimen2\font plus
\BIBentryALTinterwordstretchfactor\fontdimen3\font minus \fontdimen4\font\relax}
\providecommand{\BIBforeignlanguage}[2]{{%
\expandafter\ifx\csname l@#1\endcsname\relax
\typeout{** WARNING: IEEEtran.bst: No hyphenation pattern has been}%
\typeout{** loaded for the language `#1'. Using the pattern for}%
\typeout{** the default language instead.}%
\else
\language=\csname l@#1\endcsname
\fi
#2}}
\providecommand{\BIBdecl}{\relax}
\BIBdecl

\bibitem{gkonis2023survey}
P.~Gkonis \emph{et~al.}, ``A survey on iot-edge-cloud continuum systems: status, challenges, use cases, and open issues,'' \emph{Future Internet}, vol.~15, no.~12, 2023.

\bibitem{herrera2022joint}
J.~L. Herrera \emph{et~al.}, ``Joint optimization of response time and deployment cost in next-gen iot applications,'' \emph{{IEEE} Internet Things J.}, vol.~10, no.~5, 2022.

\bibitem{herrera2021optimizing}
------, ``Optimizing the response time in sdn-fog environments for time-strict iot applications,'' \emph{IEEE Internet of Things Journal}, vol.~8, no.~23, 2021.

\bibitem{bendechache2020simulating}
M.~Bendechache \emph{et~al.}, ``Simulating resource management across the cloud-to-thing continuum: A survey and future directions,'' \emph{Future Internet}, vol.~12, no.~6, 2020.

\bibitem{casamayor2022distributed}
V.~Casamayor~Pujol \emph{et~al.}, ``Distributed computing continuum systems--opportunities and research challenges,'' in \emph{Proc. ICSOC}, 2022.

\bibitem{pujol2023edge}
V.~C. Pujol \emph{et~al.}, ``Edge intelligence—research opportunities for distributed computing continuum systems,'' \emph{{IEEE} Internet Comput.}, vol.~27, no.~4, 2023.

\bibitem{dustdar2011principles}
S.~Dustdar \emph{et~al.}, ``Principles of elastic processes,'' \emph{{IEEE} Internet Comput.}, vol.~15, no.~5, 2011.

\bibitem{laso2022elastic}
S.~Laso \emph{et~al.}, ``Elastic data analytics for the cloud-to-things continuum,'' \emph{{IEEE} Internet Comput.}, vol.~26, no.~6, 2022.

\bibitem{murturi2022decent}
I.~Murturi and S.~Dustdar, ``Decent: A decentralized configurator for controlling elasticity in dynamic edge networks,'' \emph{ACM Trans. Internet Techn.}, vol.~22, no.~3, 2022.

\bibitem{ullah2023orchestration}
A.~Ullah \emph{et~al.}, ``Orchestration in the cloud-to-things compute continuum: taxonomy, survey and future directions,'' \emph{J. Cloud Comput.}, vol.~12, no.~1, 2023.

\bibitem{wu2020cloud}
Y.~Wu, ``Cloud-edge orchestration for the internet of things: Architecture and ai-powered data processing,'' \emph{{IEEE} Internet Things J.}, vol.~8, no.~16, 2020.

\bibitem{vcilic2021towards}
I.~{\v{C}}ili{\'c} \emph{et~al.}, ``Towards service orchestration for the cloud-to-thing continuum,'' in \emph{Proc. SpliTech}, 2021.

\bibitem{luckow2021pilot}
A.~Luckow \emph{et~al.}, ``Pilot-edge: Distributed resource management along the edge-to-cloud continuum,'' in \emph{Proc. IEEE IPDPS Workshops}, 2021.

\bibitem{masip2021managing}
X.~Masip-Bruin \emph{et~al.}, ``Managing the cloud continuum: Lessons learnt from a real fog-to-cloud deployment,'' \emph{Sensors}, vol.~21, no.~9, 2021.

\bibitem{jesus2021oppnets}
M.~Jes{\'u}s-Azabal \emph{et~al.}, ``Oppnets and rural areas: an opportunistic solution for remote communications,'' \emph{Wirel. Commun. Mob. Comput.}, vol. 2021, no.~1, 2021.

\end{thebibliography}

\vspace{12pt}
\color{red}

\end{document}